\documentclass{aa}  
\usepackage{graphicx, txfonts, hyperref, amsmath, amssymb,
amsfonts, xcolor, stfloats, booktabs, caption}
\usepackage[normalem]{ulem}

\begin{document} 
   \title{Data Representation Matters: Optimizing Machine Learning for X-ray Source Classification by Leveraging Spatio-Spectral Information}
   \titlerunning{Data Representation Matters}
   \author{         
          Maggie Lieu \inst{1}
          \and 
          Ivan Valtchanov \inst{2}
          \and
          Marguerite Pierre \inst{3} 
          \and
          Esin Gulbahar \inst{4}
          \and 
          Elias Koulouridis \inst{5, 3} 
          \and \\
          Iván Muñoz Rodríguez \inst{3}
          \and 
          Lucio Chiappetti \inst{6}
          \fnmsep}
   \institute{School of Physics \& Astronomy, University of Nottingham, University Park, Nottingham, UK, NG7 2RD \\ 
              \email{maggie.lieu@nottingham.ac.uk}
        \and 
            Telespazio UK for European Space Agency, ESAC, Camino Bajo del Castillo, s/n, Villanueva de la Ca\~nada, E-28692 Madrid, Spain 
        \and
            Université Paris-Saclay, Université Paris Cité, CEA, CNRS, AIM, 91191 Gif-sur-Yvette, France
        \and 
            Dr. Karl Remeis-Sternwarte \& ECAP, FAU Erlangen-Nürnberg, Sternwartstr. 7, Bamberg, 96049, Germany
        \and
            Institute for Astronomy \& Astrophysics, Space Applications \& Remote Sensing, National Observatory of Athens,
15236 Palaia Penteli, Greece
        \and 
            INAF, IASF Milano, via corti 12, I-20133 Milano, Italy 
             }

   \date{}

  \abstract
   { Distinguishing between faint extended X-ray sources, such as galaxy clusters, and point-like sources, primarily Active Galactic Nuclei (AGN), is a fundamental yet challenging task in X-ray surveys, crucial for cosmological studies and astrophysical source population analyses.}
   {This paper presents a benchmark comparison of convolutional neural network models for classifying simulated XMM-Newton observations of AGN and galaxy clusters, focusing on the impact of data representation and interpretability of that data.
   }
   {We systematically evaluate 1D, 2D, 2D+1D, and 3D CNNs trained on distinct data representations derived from simulated XMM-Newton observations of AGN and clusters: 1D (spectral), 2D (imaging) and 3D (event cube) data. Our experimental design leverages the physical constraints of the apparent size–redshift relation to ensure realistic source populations at fixed flux limits. We further apply 3D-GradCAM to interpret the learned features and assess the physical basis for the model's decision making to move beyond black-box classification.}
   {Our results demonstrate that increased spectral resolution enhances classifier performance using 1D data and 2D spatial information provides a strong morphological baseline, the integration of both domains yields significant diagnostic gains. We find that a 3D CNN architecture applied directly to spatio-spectral event cubes provides the most robust performance, with superior consistency across cross-validation folds. Interpretability analysis reveals that the 3D model autonomously learns to distinguish the diffuse thermal emission of the intracluster medium from the localized, non-thermal power-law signatures of AGN.}
   {The 3D CNN exploits multi-dimensional correlations that are lost in lower-dimensional projections. While the performance gain over combined 2D and 1D baselines is marginal, the 3D approach demonstrates an inherent ability to perform implicit background characterisation. By operating directly on total photon counts within event cubes, this method bypasses the need for manual background subtraction and modelling, offering a consistent, streamlined, end-to-end pipeline for the automated classification of X-ray sources in large-scale catalogues.}

   \keywords{X-rays: galaxies: clusters --
                methods: data analysis -- X-rays: general
               }

\maketitle

\section{Introduction}
The accurate classification of sources detected in X-ray surveys represents a cornerstone of modern astrophysics and cosmology \citep{clerc2023x}. Distinguishing between the two dominant populations of extragalactic X-ray emitters - Active Galactic Nuclei (AGN), powered by accretion onto supermassive black holes, and galaxy clusters, the Universe's largest gravitationally bound structures filled with hot X-ray emitting gas is critical. Precise AGN demographics are essential for understanding black hole growth, feedback mechanisms, and their co-evolution with host galaxies \citep{brandt2015cosmic, fotopoulou2016xxl}. Simultaneously, constructing clean and complete catalogues of galaxy clusters is vital for constraining cosmological parameters, testing models of structure formation, and studying the physics of the Intra-Cluster Medium (ICM). As tracers of massive dark matter halos, cluster abundance, their spatial distribution, and evolution are highly sensitive probes of cosmological parameters. Building accurate, pure cluster catalogues is essential for cluster cosmology as contamination by misclassified AGN can bias our cosmological results \citep{allen2011cosmological, pacaud2016xxl, garrel2022xxl}. The efficiency and reliability of source classification thus directly impact the scientific return of major X-ray observatories, such as XMM-Newton, Chandra, and eROSITA, as well as large-scale survey programs, such as XMM-XXL \citep{pierre2016xxl} and XCLASS \citep{koulouridis21}. 

However, despite its importance, differentiating between AGN and galaxy clusters using X-ray data alone presents significant challenges \citep{valtchanov2001comparison}. While intrinsically point-like, AGN emission is blurred by the instrumental Point Spread Function (PSF), causing them to appear spatially extended. Conversely, galaxy clusters, though intrinsically extended, may appear compact if they are distant or possess centrally concentrated emission, potentially overlapping in apparent morphology with PSF-broadened AGN. Furthermore, sources detected with low signal-to-noise ratios offer limited photons for robust morphological or spectral characterization, and source confusion in crowded fields can complicate analysis \citep{Drigga2025, Koulouridis2024}. These factors necessitate sophisticated analysis techniques to reliably separate these crucial source populations.

The XMM-Newton observatory provides a wealth of high-quality data for tackling this problem. Its European Photon Imaging Camera (EPIC) instruments function as photon-counting detectors, recording the spatial position (x, y), energy (E), and arrival time (t) for each registered X-ray event. This fundamental dataset, the "event list," contains the complete spatio-spectral information captured during an observation. From these event lists, we can then derive standard higher-level data products, such as projected 2D images (by binning events spatially) and 1D energy spectra (by binning events within specific regions according to their energy).

In X-ray astronomy, extended source classification is commonly employed to distinguish bona fide galaxy clusters from spurious detections and to quantify the reliability of the resulting samples. A widely adopted scheme defines two categories: C1, comprising sources identified as unambiguous, high-confidence cluster candidates, and C2, encompassing sources with lower statistical significance or less pronounced spatial extension \citep{pacaud2006xmm, faccioli2018xxl}. This two-tiered approach is a pragmatic way to balance the competing goals of purity (C1) and completeness (C2). The classification is typically performed by automated detection pipelines that analyse the X-ray data using statistical methods, most notably wavelet-based algorithms in conjunction with maximum-likelihood fits to model the source and determine its properties, such as flux, extent, and statistical significance. The C1 and C2 criteria (>90\% and >50\% purity levels respectively) are then applied to these measured properties to assign a class to each detected extended source.

In recent years, machine learning (ML) techniques have shown increasing promise for automating and improving source classification in X-ray astronomy. Applications have ranged from classifying sources in survey images to identifying specific spectral features \citep[e.g][]{kosiba2020multiwavelength, zhang2021classification, parker2022agn, perez2024unsupervised}. However, many studies tend to focus on applying ML algorithms to pre-processed, standard data products like images or spectra in isolation. There has been less systematic investigation into how the choice of data representation, derived from the same underlying photon event data, impacts classification performance across different ML architectures. Specifically, the potential for deep learning models to leverage the full three-dimensional (x, y, E) nature of the event list data for discriminating between point-like and extended sources has not been fully explored in a comparative context.

This paper aims to directly address this question. We conduct a systematic investigation into how different data representations impact the performance of various machine learning models for the task of classifying AGN versus galaxy clusters. Using realistic simulations tailored to XMM-Newton observations, we evaluate Convolutional Neural Networks (CNNs) specifically designed for 1D spectral data, 2D image data, and novel 3D spatio-spectral histogram data constructed from binned event lists. Our primary objective is to quantify the classification performance achieved with each data representation and ML architecture, thereby identifying approaches that most effectively utilise the rich information inherent in X-ray photon-counting data for this challenging classification problem.

We will begin by describing our simulations (\autoref{sec:data}) and models (\autoref{sec:methods}). The evaluation on test data and comparison between models is presented in \autoref{sec:results} and we discuss their implications including an interpretability analysis of the 3D model in \autoref{sec:discussion}. Finally we conclude in \autoref{sec:conclusions}. Throughout this work, we use the Planck18 cosmology model \citep{aghanim2020planck} where $H_0 = 67.66$ km Mpc$^{-1}$ s$^{-1}$ and $\Omega_M=0.31$.

\section{Data}\label{sec:data}
In these simulated 10 ks XMM-Newton observations, we perform tests under the assumption of a source with $>$50 counts in the 0.5-2 keV range. We simulate two distinct X-ray populations: galaxy clusters and AGNs. While a detection pipeline with a low threshold ($\sim$20 counts) would also naturally produce spurious sources, at 10 ks we find their contribution to be negligible within the 40–60 count range.
Consequently, our simulations focus exclusively on the distinction between real clusters and AGNs. 

Our dataset comprises simulated XMM-Newton observations categorised into three levels of complexity:

\begin{itemize}
    \item Spectral (1D): X-ray spectra representing the energy distribution of photons. These are generated using \textsc{Xspec} \citep{arnaud1999xspec}, and allows us to utilise various source models and incorporate the instrumental responses of the three XMM-Newton detectors (EPIC-pn, EPIC-MOS1, and EPIC-MOS2). This data contains the total counts and separately the combined background counts (CXB + QPB + field AGN, see \autoref{sec:background}) binned into 50 linearly-spaced bins in the energy band 0-10 keV ($\Delta$E = 0.2 keV). While classification is possible using total counts alone, this specific data representation preserves the distinction between source and background emission, reflecting the standard analytical practice of accounting for the background independently.
    \item Spatial (2D): 32×32 pixel images with 2.5 arcsecond pixel scale, encoding the spatial distribution of X-ray photons in the energy band [0.5-2 keV]. They provide information about the source morphology.
    \item Spectral-Spatial (3D): Photon event lists containing both energy and positional information for each detected photon. This represents the most complete dataset, combining spectral and spatial properties. The data are binned in three dimensions  with a spatial grid of 32$\times$32 pixels and 50 uniform energy bins spanning 0-10 keV. 
\end{itemize}

\begin{table}[]
    \centering
    \begin{tabular}{c|c|c|c|c}
         Detector & $r_c$ & $\beta$ & $A$ & $FWHM$\\
         \hline
         \hline
         EPIC-PN & 5.61 & 1.59 & - & -\\
         EPIC-MOS1 & 4.53 & 1.47 & 0.36 & 3.15 \\ 
         EPIC-MOS2 & 4.71 & 1.48 & 0.51 & 3.47\\
    \end{tabular}
    \caption{Parameters for the PSF of each XMM-Newton detector where for all detectors this is assumed to be a beta profile, with an additional a Gaussian core component for the two MOS detectors.}
    \label{tab:detector_psf}
\end{table}

To account for the instrumental blur inherent in XMM-Newton's images, we convolve the spatial distribution with an on-axis Point Spread Function (PSF) corresponding to each XMM-Newton camera. This process yields a realistic 3D dataset of photons, mimicking the spatial and spectral characteristics of an observed XMM-Newton source. The PSF tells us what an ideal point source (like an AGN) should look like, so if the observed source profile closely matches the known PSF shape and size, it's highly likely a point source (AGN). A cluster source profile is significantly more extended than the known PSF, a signature of diffuse emission of the Galaxy Cluster's ICM. In reality, while the PSF isn't constant across the  XMM-Newton field of view, we adopt a stationary on-axis PSF model for this initial study to simplify the parameter space and establish a performance baseline.

The PSF is represented by a beta profile (\autoref{eq:beta}) where the parameters of the model are listed in \autoref{tab:detector_psf}. Note that whilst the surface brightness profiles of galaxy clusters are also well approximated by the beta model, it is due to entirely different physical processes. For the EPIC-MOS detectors, there is a small, but measurable, broadening of the PSF core due to the front-illuminated design. To properly account for this, the PSF model requires an additional Gaussian core component to be added to the beta profile,
\begin{equation}
    f(x,y) =  A \exp\left(-4  \ln(2) \frac{(x-x_0)^2 + (y-y_0)^2}{{\rm FWHM}^2}\right)
\end{equation} 
where $(x_0, y_0)$ is the source centre. We use the on-axis PSF at an energy of 1 keV. To create the 3D event list, we distribute photons from the power-law spectral model (after applying the instrumental responses) randomly according to the PSF within a 2D grid defined by the input pixel size and image dimensions and the source (cluster or AGN) is randomly positioned within a circle of radius 10 arcseconds at the centre of the image which corresponds to $\pm 2$ pixels. 

We simulate the spatial distribution of the photons by distributing them randomly within a 2D grid defined by the input pixel size (2.5 arcseconds) and image dimensions ($32 \times 32$ pixels).  The resulting event list includes the following properties for each photon:
\begin{itemize}
    \item Coordinates: (xpix, ypix) within the detector
    \item Energy: ($\pi$)
    \item Instrument of detection: EPIC-pn, EPIC-MOS1, or EPIC-MOS2
    \item Source of emission: Cluster, AGN, CXB, FWC, field agn.
\end{itemize}

\subsection{Clusters}
Galaxy clusters are extended sources of diffuse X-ray emission. To simulate the X-ray spectrum from galaxy clusters, we employ an APEC model \citep{smith2001collisional}. This model describes the thermal bremsstrahlung radiation from the hot intracluster medium (ICM), incorporating key parameters such as:
\begin{itemize}
    \item Redshift ($z$): Drawn from the XXL \citep{pierre2016xxl} C1+C2 redshift distribution \citep{adami2018xxl}.
    \item Temperature ($kT$): Randomly sampled from a precomputed $z-kT$ grid. The sampling is constrained such that each simulated cluster produces a total of 50 photon counts (across all three detectors) in the [0.5–2] keV band within a 10 ks exposure, with Poissonian noise applied. Importantly, not all combinations of ($L$, $T$, $z$, $r_c$) can satisfy this constraint: for example, a hot 6 keV cluster at $z$ $\approx$ 0.1 would produce far more than 50 photons, this threshold typically selects for either intrinsically faint, low-$T$ systems at low redshift or luminous, high-$T$ systems at high redshift where cosmological dimming reduces the observed flux. This is because high-$z$ systems appear more compact and their spectra are redshifted ($T_{obs} = T / (1+z)$). This requirement thus imposes correlations between luminosity, temperature, redshift, and core radius, reducing the physically allowed parameter space. As a consequence, the spectral properties of the photons are intrinsically linked to their spatial distribution, which has direct implications for the performance of the 1D+2D and 3D methods.
    \item Metallicity ($Z$): Fixed at 0.3 solar metallicity ($Z_\odot$).
    \item Absorption: Modelled using photoelectric absorption with a hydrogen column density ($n_H$) fixed at the average value for the XXL-North field ($n_H=0.03\times 10^{22}$ cm$^{-2}$).
    \item Normalisation: Instead of being set independently, the effective normalisation is determined implicitly by the 50-photon constraint. In practice, this links back to the  luminosity-temperature scaling relation from \cite{garrel2022xxl}, ensuring the simulated X-ray luminosity ($L_X$) within a specified energy band matches observational constraints while respecting the count-limited selection.
\end{itemize}

In addition, we pre-compute the cluster core radii $r_c$ using the \cite{garrel2022xxl} scaling relations,
\begin{equation}
    \frac{M_{500,WL}}{2.6 \times 10^{13}  M_\odot h^{-1}} = \left(\frac{T_{300 {\rm\, kpc}}}{1 {\rm\, keV}
    }\right)^{1.67} E(z)^{-1}
\end{equation}
\begin{equation}
    r_{500} = \sqrt[3]{\frac{3 M_{500,WL}}{4 \pi 500 \rho_{crit}(z)}}
\end{equation}
where $M_{500,WL}$ is the weak-lensing mass estimate within $r_{500}$, $r_c = 0.15 r_{500}$ is the core radius, $T_{300}$ kpc is the cluster X-ray temperature measured inside 300 kpc, $\rho_{crit}(z)$ is the critical density of the Universe at redshift $z$, and $E(z)$ is the redshift evolution of the Hubble parameter, $E(z) \equiv H(z)/H_0$.

To generate spatially-resolved event lists from the APEC spectrum, we model the cluster's X-ray surface brightness using a beta profile (also known as a King profile or Moffat profile) \citep{king1966structure}. This profile provides an analytic approximation to the isothermal sphere and is characterised by a core radius ($r_c$) and a slope parameter ($\beta$). We adopt a fixed $\beta$ value of 1.5 and utilise the precomputed core radii for each cluster with an uncertainty of 0.2 to mimic scaling relations uncertainties.
The 2D beta profile is given by:
\begin{equation}\label{eq:beta}
f(x,y) = \left( 1+ \frac{(x-x_0)^2 + (y-y_0)^2}{r_c^2} \right)^{-\beta}.
\end{equation}

\subsection{AGN}\label{sec:AGN}
For AGN spectral emission, we adopt a power-law model modified by Galactic absorption, incorporating the following parameters:
\begin{itemize}
    \item Power-law index $\Gamma$:  The primary continuum slope is randomly sampled from a uniform distribution between $1.5$ and $2.0$. This range is chosen to represent the intrinsic diversity of Seyfert galaxies and quasars \citep[e.g.][]{reeves2000x, porquet2004xmm}. While high-redshift or high-Eddington ratio sources can exhibit $\Gamma > 2.0$, the $1.5–2.0$ range captures the dominant population of AGN while maintaining consistency with XMM-Newton spectral extraction defaults \citep{mateos2005xmm}.
    \item Absorption: Modelled using the \texttt{tbabs} model in \textsc{Xspec}, representing interstellar medium (ISM) grain absorption. We consider column density ($n_H$) randomly drawn from a truncated log-normal distribution with limits 0.01 to 10.0 $\times 10^{22}$ cm$^{-2}$. This ensures the inclusion of intrinsically obscured Type 2 AGN, which exhibit significantly harder spectra than the diffuse, soft thermal emission of clusters \citep{tozzi2006x, ueda2014toward}.
\end{itemize}
To normalise the AGN spectrum, we fix the number of counts in the EPIC-pn detector for a given exposure time to be 32 and with the differences in effective area with MOS detectors this normalisation ensures the total pn+MOS1+MOS2 counts of $\sim 50$ with an associated Poissonian error. We then calculate the corresponding flux in the [0.5, 2.0] keV band. This calculated flux, which is on the order of $5 \times 10^{-15}$ erg s$^{-1}$ cm$^{-2}$ for our assumed spectral model, and is well above the nominal AGN detection threshold for the XXL survey, which is approximately $10^{-15}$ erg s$^{=1}$ cm$^{-2}$.

\subsection{Background}\label{sec:background}
Our simulations incorporate realistic background components that are crucial for accurately representing XMM-Newton observations. These background components include the Cosmic X-ray background (CXB), the Quiescent Particle Background (QPB) and field AGN (FAGN). These components contribute average count rates in the 0.5-2 keV band of $6.21 \times 10^{-6}$, $4.38 \times 10^{-6}$, and $6.47 \times 10^{-7}$ counts/s/pixel, respectively, and are consistent with other studies \citep{valotti2018cosmological}.

\subsubsection{Cosmic X-ray background}\label{sec:CXB}
The CXB is modelled in XSPEC following the guidelines in the ESAS Cookbook\footnote{\url{https://heasarc.gsfc.nasa.gov/docs/xmm/esas/cookbook/xmm-esas.html}} \citep{snowden2014cookbook}. This approach decomposes the CXB into four distinct components, corresponding to the emission from the Local Hot Bubble and heliosphere, the emission from the cooler galactic halo, emission from the hotter regions of the galactic halo and intergalactic medium and emission from the unresolved background of cosmological sources (primarily active galaxies and quasars).  We randomly sample cosmic background spectrum for each simulation.

\subsubsection{Quiescent particle background}\label{sec:FWC}
The QPB arises from the interaction of high-energy particles with the XMM-Newton detectors. These particles can be cosmic rays or particles trapped in the Earth's radiation belts. The QPB is characterised by a complex spectral shape and spatial distribution that varies across the detector and over time. 

To model the QPB XMM-Newton observations include dedicated exposures with the filter wheel closed (FWC), blocking astrophysical X-rays and allowing direct measurement of the QPB. We specifically use the FWC data provided in the Current Calibration Files (CCF)\footnote{Available from \url{https://www.cosmos.esa.int/web/XMM-Newton/current-calibration-files}} to characterise the QPB spectrum and its spatial variation.  

For each Observation ID (OBSID), we extract event lists from the corresponding FWC CCF files and create subsets representing 10 ks of exposure time using \texttt{XMM-SAS v21}. These subsets are stored for use in our simulations. To incorporate the QPB into a simulated observation, we randomly select an OBSID from a pool of available FWC observations to allow for different levels of background. The backgrounds are different for each instrument and also vary with the solar cycle. The events from the selected 10 ks FWC observation are then directly added to the simulated 3D event list, which contains the list of photons from the astrophysical source and the cosmic X-ray background.

\subsubsection{Other point sources}\label{sec:fagn}
Besides the central source, observations may also contain point sources such as field AGN residing in the background or AGNs located within the galaxy cluster itself, often associated with member galaxies. The flux from these sources can blend with the emission from the central source, potentially leading to misidentification of the central source. To take this into account, we use the phenomenological calculation of logN-logS\footnote{\url{http://www.bo.astro.it/~gilli/counts.html}} as detailed in \cite{gilli2007synthesis} to estimate the number density of AGNs as a function of flux, redshift, and absorption. The input parameters for this calculation are:
\begin{itemize}
    \item Energy band: [0.5, 2] keV
 \item Flux range: $[5\times10^{-15},10^{-11}] \rm\ erg\ s^{-1} cm^{-2}$
 \item Absorption $\ln(n_H)$:$ [20, 21] \rm cm^{-2}$
 \item Redshift ($z$): [0.1, 4]
 \item Luminosity $\ln(L_X)$: [40, 45] erg s$^{-1}$
\end{itemize}
This calculation provides us with the total number of AGNs per square degree within the specified parameter ranges. To allow for representative field AGN sample, we scale the predicted numbers for a 25 square degree field to match the XXL coverage \citep{pierre2016xxl}, randomly placing the AGNs and drawing fluxes from the logN-logS.

For each simulated source (either the central cluster or an AGN), we randomly throw a 80x80 arcsec box within the 25 sq. deg. field with field AGNs and if there is an AGN within this box, it is added to the simulated image. The spectral and spatial modelling of these additional AGNs follows the same approach as described in \autoref{sec:AGN}. We do not discard cases when the field AGN centre lies within the central 10 arcsec thus blend with the central source.

\begin{figure*}
    \centering
    \includegraphics[width=0.60\linewidth]{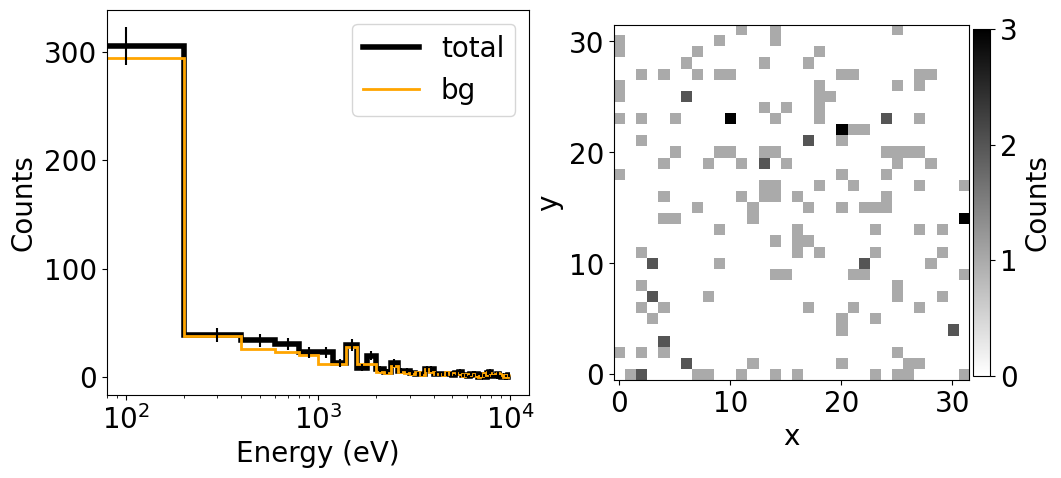}
    \includegraphics[width=0.32\linewidth]{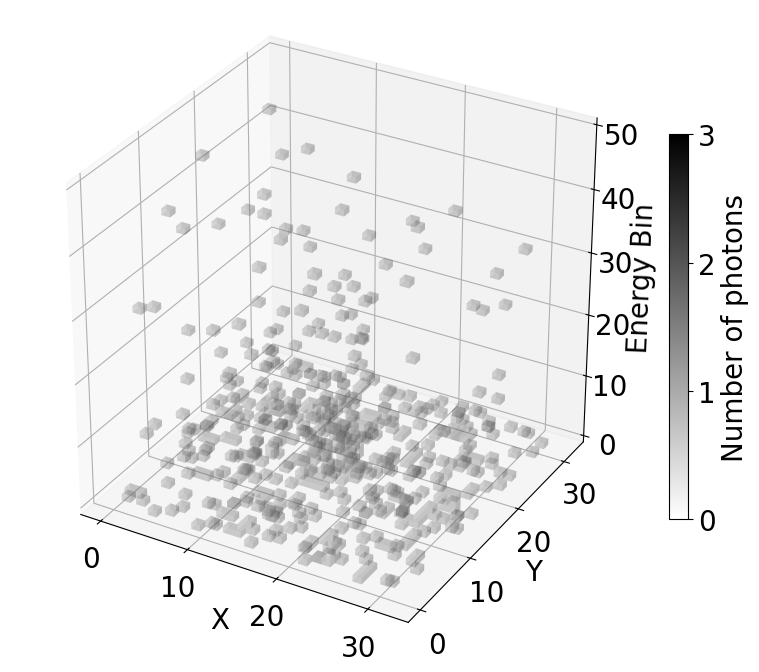}
    \caption{From left to right, an example of 1D, 2D and 3D data of a cluster simulation. The spectra (1D data) show both the total counts and the background consisting of the filter wheel closed counts and the cosmic X-ray background. Note that in some bins this is higher than the total counts due to Poisson sampling on the total counts. The error bars on the total counts show the Poisson uncertainty. There are no field AGN in this pointing.}
    \label{fig:excluster}
\end{figure*}

\begin{figure*}
    \centering
    \includegraphics[width=0.60\linewidth]{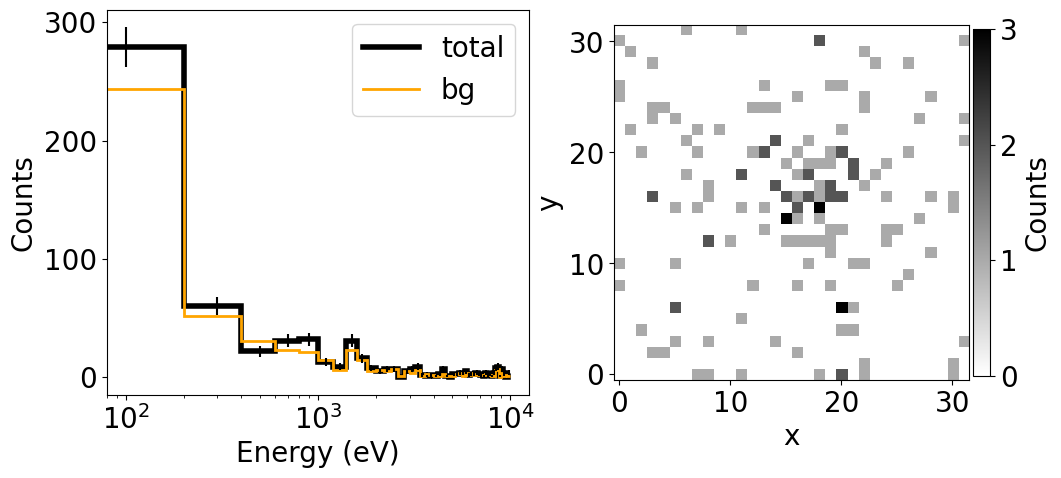}
    \includegraphics[width=0.32\linewidth]{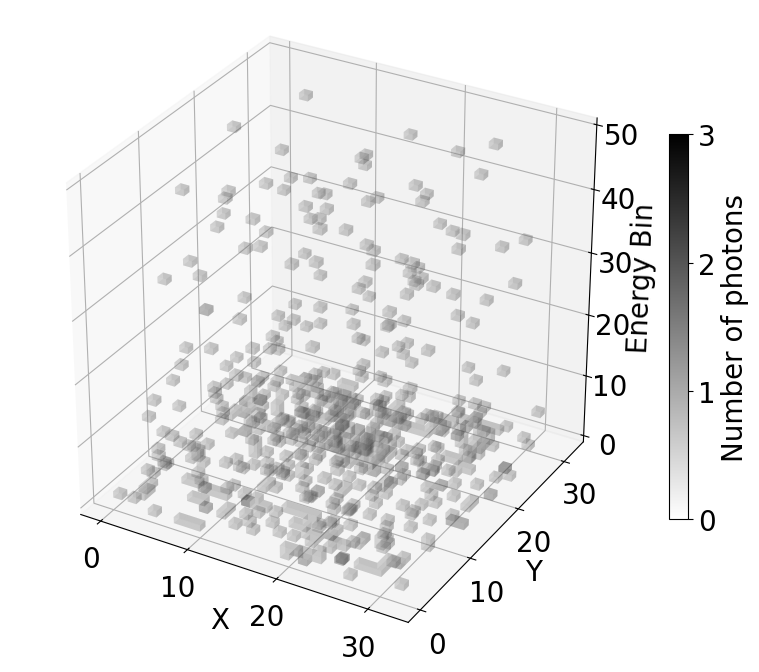}
    \caption{From left to right, an example of 1D, 2D and 3D data of an AGN simulation. The spectra (1D data) show both the total counts and the background consisting of the filter wheel closed counts and the cosmic X-ray background. Note that in some bins this is higher than the total counts due to Poisson sampling on the total counts. The error bars on the total counts show the Poisson uncertainty. There are no field AGN in this pointing.} 
    \label{fig:exagn}
\end{figure*}

\section{Method}\label{sec:methods}
The total dataset consists of 1600 clusters and 3440 AGN for which we have the raw and binned photon event lists (i.e. position and energy of hits) (3D data),  images (2D data) and spectra (1D data). For the clusters, the data is split into 80-10-10 training, validation, test datasets. For the AGN and spurious sources, the training and validation sets are matched in number to those of the clusters to ensure balanced learning.
However, the test set composition reflects realistic source abundances, with approximately 7\% (160) clusters, and 93\% (2000) AGN consistent with expectations from the findings of the XXL survey \citep{adami2018xxl, chiappetti2018xxl}.

\subsection{Convolutional neural networks}
The primary challenge lies in distinguishing between competing sources, such as AGN and galaxy clusters, which can exhibit similar observational signatures in X-ray surveys. CNNs have emerged as a powerful tool for automated feature extraction from complex data, using kernels that slide along the data to detect local patterns \citep{lecun2015deep}. The utility of image-based convolution is well-established in astronomy; for example, the FIELD software for Subaru/FMOS uses 2D convolution and signal-to-noise ($S/N$) mapping to automatically detect emission lines while filtering out cosmic rays and instrumental artifacts \citep{tonegawa2015field}. Their hierarchical architecture allows them to learn increasingly complex spatial and temporal hierarchies, by combining simple local patterns, making them highly effective for classification. For a fair and direct comparison across all data types, we employ a unified CNN-based approach, avoiding a mixture of different methods that could introduce model-specific biases. Here we propose a novel approach that leverages the full-event data from X-ray observatories by applying specialised CNNs to each of the 1D, 2D, 2D+1D and 3D data to determine which data representation is most informative for classification. This multi-modal strategy is motivated by the distinct physical properties and spatial distributions of AGN and galaxy clusters, which are captured uniquely within each data representation. While model parameters are updated exclusively using the training data, we employ a 9-fold cross-validation strategy to evaluate performance across the entire development set. This iterative process ensures the model's robustness and reduces the risk of overfitting to a specific data split. We train each model for a maximum of 100 epochs, monitoring the progress with the validation data and implementing early stopping to prevent overfitting and ensure generalisation to new, unseen data. The models are trained in batch sizes of 32 and the Adam optimizer with an initial learning rate of 0.001, which decays by a factor of 0.1 when learning stagnates. As this is a binary classification problem (cluster vs. AGN), the final layer in all the models are a single neuron dense layer and the loss function being minimised is binary cross entropy. This is configured to output raw logits to enhance numerical stability during backpropagation by avoiding the squashing effects of an intermediate sigmoid activation. All other layers use the ReLU activation function for non-linearity. 

\subsection{1D Model}\label{sec:1d}
The 1D model focuses solely on the spectral information of the total counts and combined background counts. These data are input as separate two 50-linear bin vectors; to account for the statistical nature of photon detection, we augment the training data by sampling from a Poisson distribution based on the total counts, ensuring the model is robust to realistic observational noise and ensures that the true source counts remains unknown. The model used is a 1D CNN (\autoref{tab:1d_cnn_model_summary}), that is optimized for sequential data that has a single spatial or temporal dimension such as energy spectra. While effective for pattern recognition, 1D CNNs require sufficient input dimensionality to function correctly; standard 5-bin spectra are unsuitable, as the data dimensions become comparable to the kernel size. By utilising 50 linearly-spaced bins (motivated in \autoref{sec:mlp}), we provide a high-resolution feature space that allows the 1D kernels to extract nuanced spectral features and gradients without encountering the edge-effect artifacts or information loss inherent in lower-dimensional inputs.

\subsection{2D Model}\label{sec:2d}
The 2D model is designed to exploit the spatial information present in the X-ray images. To analyse these, we employ a two-dimensional convolutional neural network (2D CNN), which applies 2D learnable filters across the image plane to capture hierarchical spatial patterns. CNNs have already proven effective in X-ray astronomy, including galaxy cluster classification tasks \citep{sweere2022deep, kosiba2020multiwavelength}.

Through convolution and pooling operations, the network learns features such as edges, textures, and morphological structures at multiple scales, while reducing dimensionality, introducing translational invariance \citep{lieu2019detecting}. The architecture is summarised in \autoref{tab:2d_model_summary}.

\begin{table*}[h]
\centering
\small
\caption{Model Performance Metrics.}
\label{tab:results}
\begin{tabular}{lccccccccccc}
\toprule
Model & TP & FP & FN & TN & Accuracy & Precision & Recall & F1 & MCC & ROC AUC & PR AUC \\
\midrule
1D & 96 & 98 & 63 & 1901 & 0.9251 & 0.5012 & 0.6049 & 0.5451 & 0.5093 & 0.9151 & 0.6052 \\
   & $\pm$ 8.2 & $\pm$ 22.4 & $\pm$ 8.2 & $\pm$ 22.4 & $\pm$ 0.0077 & $\pm$ 0.0489 & $\pm$ 0.0510 & $\pm$ 0.0228 & $\pm$ 0.0266 & $\pm$ 0.0085 & $\pm$ 0.0293 \\
\midrule
2D & 158 & 1 & 1 & 1998 & 0.9983 & 0.9890 & 0.9882 & 0.9885 & 0.9877 & 0.9989 & 0.9977 \\
   & $\pm$ 1.7 & $\pm$ 1.9 & $\pm$ 1.7 & $\pm$ 1.9 & $\pm$ 0.0010 & $\pm$ 0.0118 & $\pm$ 0.0104 & $\pm$ 0.0071 & $\pm$ 0.0076 & $\pm$ 0.0016 & $\pm$ 0.0023 \\
\midrule
2D+1D & 158 & 0 & 2 & 1999 & 0.9989 & 0.9972 & 0.9875 & 0.9922 & 0.9917 & \textbf{0.9999} & \textbf{0.9992} \\
      & $\pm$ 3.6 & $\pm$ 0.7 & $\pm$ 3.6 & $\pm$ 0.7 & $\pm$ 0.0016 & $\pm$ 0.0043 & $\pm$ 0.0222 & $\pm$ 0.0114 & $\pm$ 0.0120 & \textbf{$\pm$ 0.0002} & \textbf{$\pm$ 0.0011} \\
\midrule
3D & \textbf{158} & \textbf{0} & \textbf{1} & \textbf{1999} & \textbf{0.9993} & \textbf{0.9986} & \textbf{0.9924} & \textbf{0.9955} & \textbf{0.9951} & 0.9970 & 0.9958 \\
   & \textbf{$\pm$ 0.4} & \textbf{$\pm$ 0.4} & \textbf{$\pm$ 0.4} & \textbf{$\pm$ 0.4} & \textbf{$\pm$ 0.0002} & \textbf{$\pm$ 0.0026} & \textbf{$\pm$ 0.0026} & \textbf{$\pm$ 0.0016} & \textbf{$\pm$ 0.0017} & $\pm$ 0.0024 & $\pm$ 0.0021 \\
\bottomrule
\end{tabular}
\tablefoot{This table summarises the classification results on the test dataset for the four candidate models. Values represent the mean $\pm$ standard deviation across all 9-folds. To facilitate comparison in a high-performance environment, the best-performing model for each metric is highlighted in bold. For error-based metrics (FP and FN), the minimum value is highlighted; for all other metrics, the maximum value is highlighted. While the degree of precision may exceed the practical limits of the uncertainties; they are maintained here to ensure a granular comparison across all.}
\end{table*}

\begin{table}[]
    \centering
    \caption{Model training statistics.}
    \begin{tabular}{l|c|c|c|c|c|c}
        Model & 1D & 2D & 2D+1D & 3D\\
        \hline
        Training time & 2 ms & 3 ms & 5 ms & 12 ms \\
        Epochs & 28 & 34 & 27 & 12 
    \end{tabular}
    \tablefoot{The average per-step training times of the various models on two NVIDIA RTX A6000, and the number of epochs.}
    \label{tab:training_time}
\end{table}
\subsection{2D+1D Model}\label{sec:2dp1d}
Our aim is to evaluate classification performance using spatial information alone and compare it against models that incorporate spectral or joint spectral–spatial features. This provides a direct measure of the relative importance of spatial and spectral information in distinguishing galaxy clusters from AGNs. Spatial and spectral data provide complementary information about the X-ray sources. Combining them can potentially lead to improved classification accuracy. Incorporating both types of data, may also make the model more robust to noise and variations in individual data modalities. To leverage both the spatial and spectral information available in our dataset, we construct a 2D+1D model that combines the strengths of the 2D and 1D CNN, enabling the network to capture the interplay between spatial morphology and spectral signatures.

The 2D+1D architecture involves two parallel branches, the 2D branch extracts spatial features from the 32$\times$32 pixel images following the architecture described in the 2D model. The 1D branch extracts spectral features from the spectral data and follows the architecture described for the 1D model. Their outputs are concatenated to form a unified feature representation, which is then passed through a dense layer for higher-level feature integration and a final dense layer for classification. The detailed architecture of the model, including the connections between layers, is summarised in \autoref{tab:2dp1d_model_summary}.

\subsection{3D Model}
The 3D model utilises the 3D data containing the complete raw event list data to provide the most comprehensive information about each X-ray observation.  This data includes the position ($x_{pix}, y_{pix}$) and energy ($\pi$) of individual photons detected by the XMM-Newton telescope binned into the 32$\times$32$\times$50 voxels. 

The architecture of the 3D model is detailed in \autoref{tab:3d_model_summary}. This architecture allows the model to learn complex patterns and interactions between photon positions and energies, capturing the combined spatial and spectral information inherent in the raw event data. This potentially improves accuracy and robustness compared to models relying on reduced data representations.

\section{Results}\label{sec:results}
We evaluate the 4 models (\autoref{tab:results}) on the imbalanced test dataset with the standard metrics: accuracy, precision, recall and F1-Score. The accuracy represents the overall proportion of correctly classified sources. Precision is a measure of the purity of the positive class sample namely the clusters. Recall is a measure of the completeness of the positive sample and F1-score is a harmonic mean between the two. In addition we compute the Matthews correlation coefficient (MCC), which is  single-number metric for binary classification, typically reserved for imbalanced datasets. Unlike F1, it takes all four quadrants of the confusion matrix into account,
\begin{equation}
    MCC = \frac{TP \times TN - FP \times FN }{\sqrt{(TP+FN)(TN+FP)(TN+FP)(TN+FN)}},
\end{equation}
Where TP, FP, TN and FN are the number of true positive, false positive, true negative and false negative predictions respectively, with clusters being the positive class. We present the receiver operating characteristic (ROC) and precision-recall (PR) curves (\autoref{fig:ROC_PR}) and compute the area under the curve (AUC). These metrics evaluate the performance whilst taking into account the different thresholds of the classification, with the PR curve offering superior insight for the imbalanced dataset by focusing on the minority class (clusters) and ignoring true negatives. Recall that there are an order $10 \times$ more AGN than clusters. 

\begin{table}[]
    \centering
    \caption{Model comparison via diagnostic odds ratio.}
    \begin{tabular}{l|c|c|c}
        Model & $\overline{L}$ & 95\% CI & p-value \\
        \hline 
        1D & 3.41 & [3.27, 3.55] & 1.12$\times 10^{-11}$\\
        2D & 11.55 & [10.78, 12.32] & 0.016\\
        2D+1D & 12.54 & [11.76, 13.32] & 0.867\\
        3D & 12.60 & [12.28, 12.92] & -
    \end{tabular}
    \tablefoot{This is presented as the mean log diagnostic odds ratio ($\overline{L}$) over the 9-cross validation runs. The 95\% confidence intervals are similarly calculated over the log-normal transformation. P-values represent the results of a paired t-test on the log odds ratio relative to the 3D model. }
    \label{tab:significance}
\end{table}

Our results indicate across all metrics the 3D model outperforms the other models, except in the case of AUC, where the the 2D+1D (factorised) model excels. To examine the statistical significance, the log diagnostic odds ratios (DOR) are compared across the 9 cross-validation folds (\autoref{tab:significance}) and a paired t-test is used to compute the p-values. While the 3D architecture achieves the highest mean log-odds ratio, compared to the 2D+1D model, the difference did not reach statistical significance ($p$=0.87). None the less, the 3D model demonstrated significantly higher precision in its performance with a standard error (0.16) less than half of the 2D+1D baseline (0.38). The p-value with respect to the 2D model suggests the higher DOR is statistically significant.

\section{Discussion}\label{sec:discussion}

The discrepancy between the 2D+1D and 3D model performance occurs because AUC measures a fundamentally different aspect of a model’s performance. AUC measures how well the model separates classes across all possible thresholds. A higher AUC for the 2D+1D model means it is better at assigning a higher probability to a positive sample (cluster) than a negative sample (AGN). It is better at ranking the probabilities. Whereas the other metrics (accuracy, F1 etc.) are computed at a fixed threshold (0.5) and show performance better on the final classification. If we were to calibrate the threshold, the 2D+1D model would likely outperform the 3D model in those metrics. The 2D+1D architecture factorises a 3D convolution into a spatial 2D filter and a temporal 1D filter. This factorisation introduces more non-linearities which helps the model represent more complex functions, leading to better feature discrimination, which is what AUC captures, whereas the full 3D convolutions learn spatial and temporal features simultaneously, leading to higher certainty and sharpness in the final predictions. 

This is also reflected in the training times (\autoref{tab:training_time}). The 3D architecture had faster convergence despite having more parameters and higher computational cost per step, likely due to this simultaneous spatial-temporal modelling. It is able to find an optimal solution more quickly. The 2D+1D model has a bottleneck and factorisation overhead in joining the two parts of the model. The parallel branches increases the computational graph complexity as the model needs to perform forward and backward passes through both branches simultaneously. The model would need to synchronise the gradients from both branches before updating the weights and this synchronisation step can introduce delays, particularly if one branch takes significantly longer to process than the other. Additionally with parallel branches, there's more data movement between the CPU and GPU (or between different parts of the GPU). This data transfer can be a bottleneck, especially if the data is large or if there are frequent transfers. We note however that the differences in performance is minuscule and the inference times are comparable for all models (<1ms per step). In practice, the simpler method would be preferred.

\begin{figure}
    \centering
    \includegraphics[width=\linewidth]{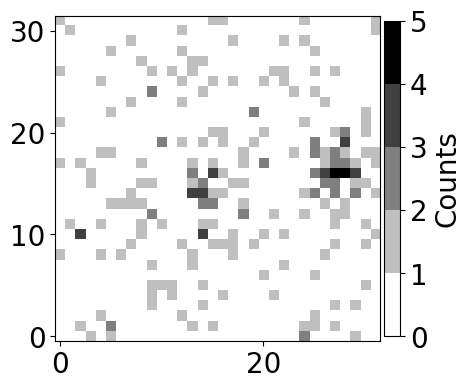}
    \caption{Image of the single false positive detection found in two of the nine 3D model runs. }
    \label{fig:FPs}
\end{figure}

\begin{figure}
    \centering
    \includegraphics[width=\linewidth]{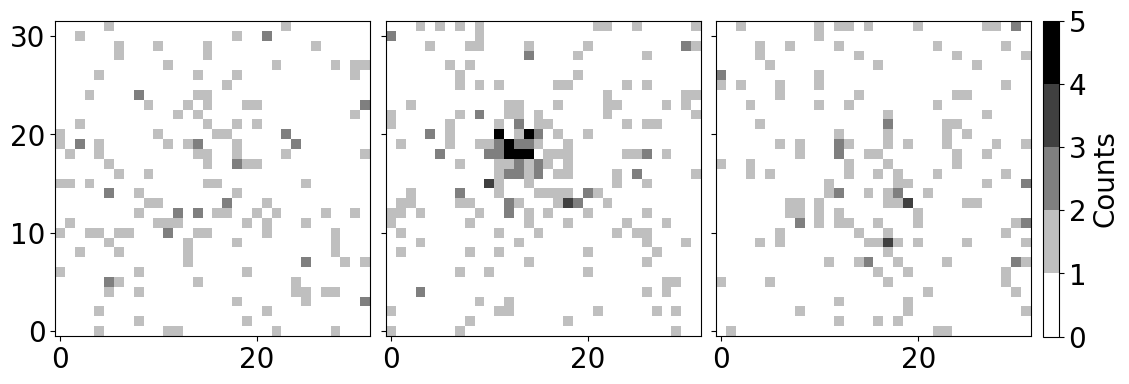}
    \caption{Images of the false negative cluster detections in the 3D models. The central cluster is contaminated with field AGN and this detection is a false negative in all nine cross-validation runs. As previously said, the central object has far more than 50 photons, hence would not be selected to be analysed by the CNN trained for 50. The other two are high redshift and  small core radii. They only occur in one of the nine runs. }
    \label{fig:FNs}
\end{figure}

The superior performance of the 3D and 2D+1D models suggests that combining spatial and spectral information is indeed beneficial for distinguishing between galaxy clusters and AGNs. These models likely capture the complementary nature of these two data modalities, leading to more robust and accurate classifications.

The 2D model outperforming the 1D model is expected, as spatial information plays a crucial role in differentiating extended sources like galaxy clusters from point sources like AGNs. This result underscores the importance of incorporating spatial features in the classification process.

The marginal improvement of the 2D+1D model over the 2D model suggests that while spectral information can enhance the classification, the spatial features extracted by the CNN are the primary drivers of performance. This observation could be valuable for designing future models, potentially focusing on optimising the CNN architecture or exploring more sophisticated ways to combine spatial and spectral information.

\begin{figure}
    \centering
    \includegraphics[width=\linewidth]{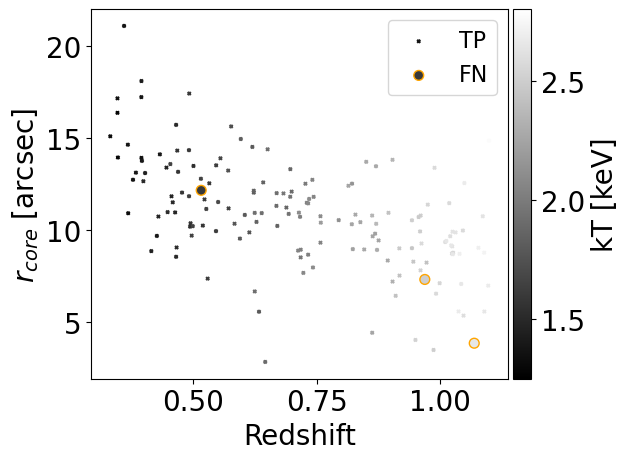}
    \caption{The distribtion of clusters in the test dataset in core radius, redshift and temperature space. Clusters missed by the 3D network are shown with enlarged markers.}
    \label{fig:fnsrcore}
\end{figure}

Focusing on exclusively on the 3D model, across the 9 cross-validation runs, only one AGN was misclassified as a cluster (\autoref{fig:FPs}). This specific source exhibits a low pn-flux 5.129$\times 10^{-15}$ erg s$^{-1}$ cm$^{-2}$ and is contaminated by a field AGN. The combined emission from the cluster and field AGN results in an integrated count exceeding the 50-photon threshold which is not representative of the isolated low-count regime the model is designed to evaluate. However, simulation data (Muñoz et al., in prep.) highlights the strength of this architecture: out of some 100 input clusters that have been detected with 40-60 photons, the traditional classification method \citep{pacaud2006xmm} characterises  only 10\% of them as clusters\footnote{we note however that the simulations were run with full modelling of the PSF off-axis dependence}. In contrast, the 3D CNN successfully recovers the majority of these faint objects. This demonstrates that while contamination remains a challenge, the neural network significantly expands the detectable population of clusters at the sensitivity limit compared to traditional analytic approaches. Conversely, three clusters were missed by the 3D model (\autoref{fig:FNs}); these sources are characterised by notably high redshift and small core radii (\autoref{fig:fnsrcore}) and the lower redshift cluster is contaminated by a central AGN.

\subsection{Interpretability analysis}
Interpretability is crucial for assessing the physical basis of a model's decision making process, and such analysis is already emerging as the standard for trustworthy AI in science and astronomy \citep{ntampaka2022importance, huang2025interpretable}. To interpret the decision making process of the 3D model, we apply Gradient-weighted Class Activation Mapping \citep[Grad-CAM][]{selvaraju2017grad} to visualise the regions of the input data most influential to the model's predictions \citep[see][for a review]{lieu2025comprehensive}. We perform the analysis using the best model from the 9 cross-validation models. Specifically, we calculate the gradients of the target class score with respect to the feature maps of the final convolutional layer. These gradients are globally pooled to serve as importance weights which are then used to produce a weighted sum of the feature maps. After applying a ReLU activation function to produce an attention map, we rescale to the original input dimensions (32$\times$32$\times$50). Given the volumetric nature of our data, we condense this 3D attention map into spectral and spatial components by taking a maximum intensity projection (MIP) along the respective axes. This approach allows us to isolate the peak optimal discriminators. These are the specific energy bins and spatial coordinates that most strongly drive the classification of galaxy clusters and AGN. 

\begin{figure}
    \centering
    \includegraphics[width=\linewidth]{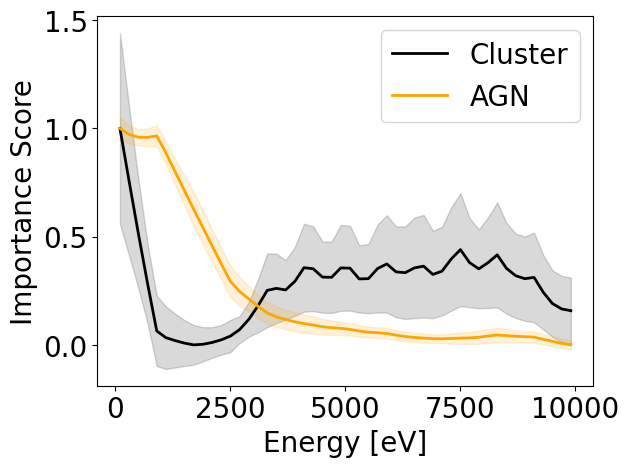}
    \caption{Mean normalised Grad-CAM importance profiles for the correctly classified galaxy clusters (N=159) from the test set, showing the model's spectral attention across the 0-10 keV energy range.}
    \label{fig:gradcam1d}
\end{figure}

\begin{figure}
    \centering
    \includegraphics[width=\linewidth]{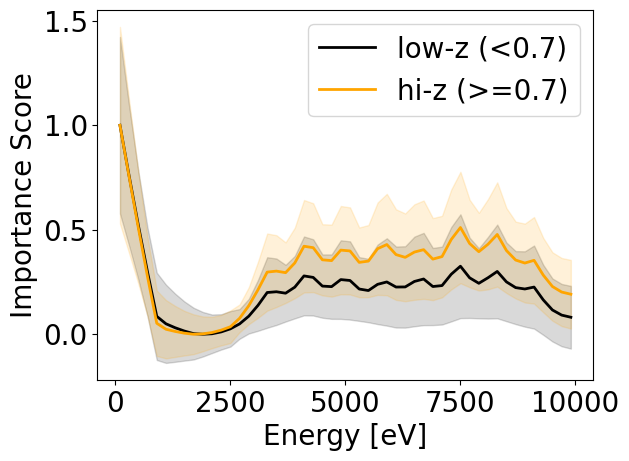}
    \caption{Mean normalised Grad-CAM importance profiles for all galaxy clusters (N=1600) from the entire data set split by redshift, showing the model's spectral attention across the 0-10 keV energy range.}
    \label{fig:gradcam1dz}
\end{figure}

\begin{figure}
    \centering
    \includegraphics[width=\linewidth]{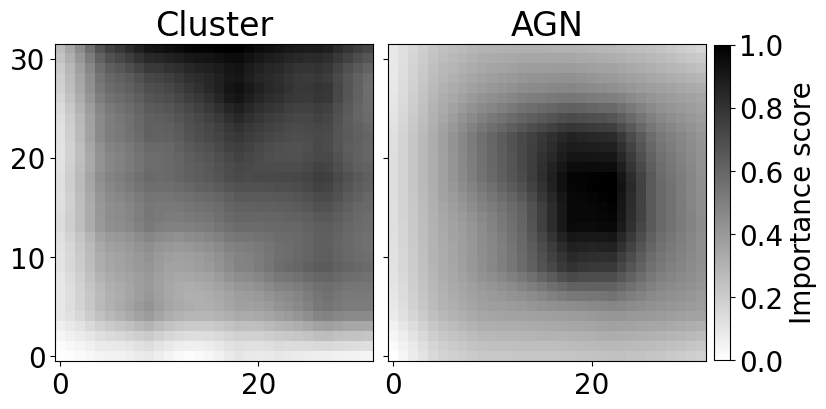}
    \caption{Mean normalised spatial Grad-CAM importance maps for correctly classified galaxy clusters and AGN in the test set. The axis correspond to pixels in the 32$\times$32 input space.}
    \label{fig:gradcam2d}
\end{figure}

The average spectral and spatial attention profiles are created by averaging over all correctly classified test data. In the spectral domain (\autoref{fig:gradcam1d}), the AGN profile follows a single-peaked distribution primarily weighted towards the low energies (<2.5 keV), consistent with the soft X-ray power-law where AGN signal-to-noise ratio is highest. In contrast the clusters exhibit a bimodal importance distribution, an initial peak in the soft X-ray region (< 1keV) where thermal emission is most prominent followed by a sustained importance across the hard band. Interestingly the drop in importance around 2 keV coincides with a known feature in the XMM-Newton spectral sensitivity (effective area) caused by the Gold (Au) M-shell absorption edge ($\sim$2.2 keV) and the Silicon K-shell Absorption Edge ($\sim$1.8 keV).

To quantify this we split the profiles into soft (0-2 keV), medium (2-6 keV) and hard (6-10 keV) bands and calculate the area under the attention curves. The resulting attention fractions for clusters [0.19, 0.35, 0.46] and AGN [0.70, 0.25, 0.05] highlight a distinct divergence in the model's classification logic, for clusters, nearly 50\% weighting in the hard band where the source continuum is typically fainter suggests the model is not only searching for source brightness which peaks in the soft band but instead uses the stationary instrumental features in the hard tail as a reference. Conversely for AGN, the 70\% dominance in the soft band suggests the model identifies sources from their high-signal-to-noise ratio and characteristic point-like PSF in the softest channels where non-thermal power law emission is most prominent.

The redshift-binned profiles (\autoref{fig:gradcam1dz}) reveal that the soft peak is consistent across redshift, while hard-band importance is systematically elevated for high-$z$ clusters across $\sim$3–10 keV. This is consistent with the model compensating for reduced soft-band source-to-background contrast at higher redshift by placing greater reliance on stationary hard-band features. These may correspond to the limitations of the background modelling: the small number of unique FWC observations may result in repeated noise patterns across the data set. Consequently, the 'wiggles' observed in the importance profiles may partially reflect insufficient sampling of the instrumental background rather than generalised instrumental features, suggesting that the model may be learning specific noise realisations unique to this limited dataset.

In the spatial domain (\autoref{fig:gradcam2d}), the average AGN GradCAM map reveals a tight, centrally concentrated peak consistent with the morphology of point-sources, where the high intensity flux concentration serves as a primary diagnostic feature. The cluster attention map is notably more broad, capturing the diffuse and extended nature of the intracluster medium with elevated importance extended to the image periphery. 

\begin{figure}
    \centering
    \includegraphics[width=\linewidth]{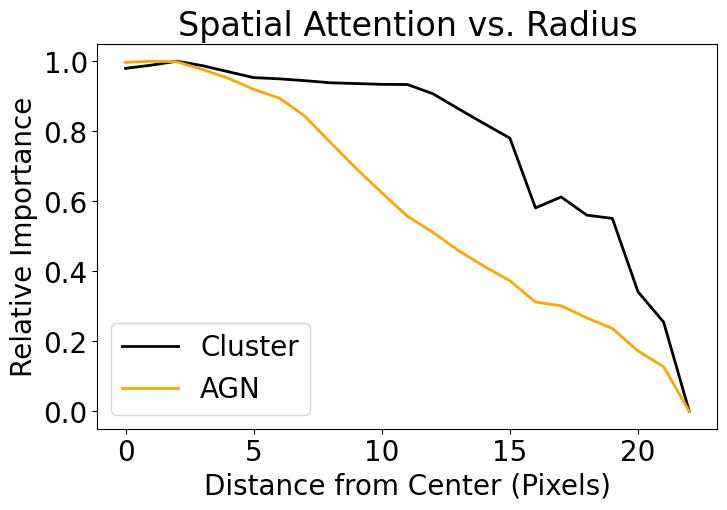}
    \caption{Radial profiles of spatial importance for clusters and AGN.}
    \label{fig:radial_att}
\end{figure}

To quantify the spatial importance, we generate radial importance profiles (\autoref{fig:radial_att}). The cluster profile has maximum importance at the centre and maintains a high-plateau out to 12 pixels before a sharp decline. Since the spatial inputs are 32 $\times$ 32, this dominates the entire image and confirms the model recognises the cluster as a large, centralised structure composed of both a core and a significant spatial envelope. The AGN importance also peaks at the centre but drops smoothly, consistent with the decay of a Gaussian PSF profile. This divergence demonstrates that while the AGN classification is driven by a localised peak, the cluster classification relies on a spatially integrated volume of information.

\begin{figure}
    \centering
    \includegraphics[width=1.1\linewidth]{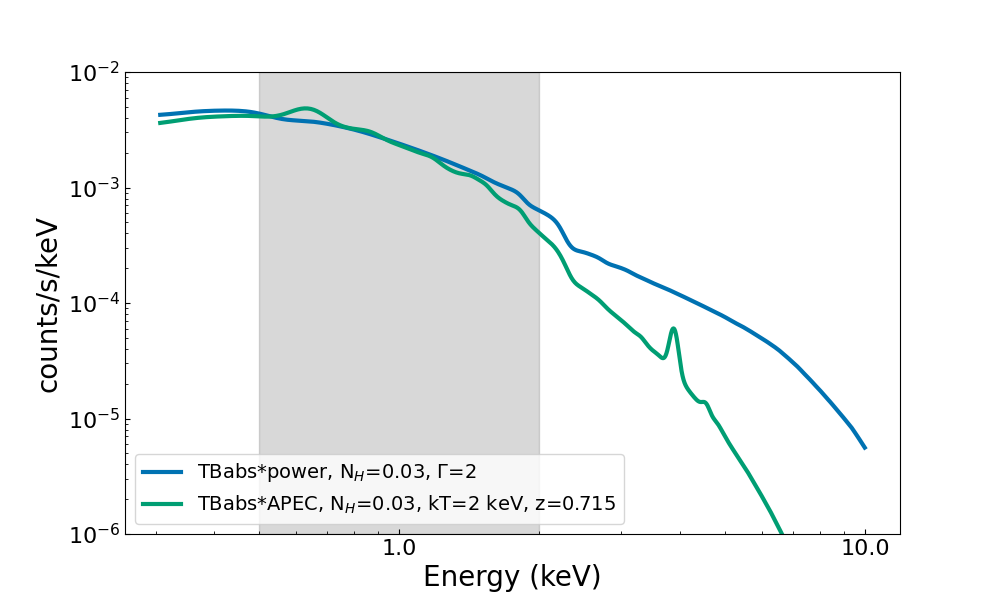}
    \caption{A comparison of the X-ray emission from optically thin plasma modelled with APEC \citep{smith2001collisional} and a power law model, illustrating the similarity between the profiles with similar flux levels in [0.5,2] keV (shaded area). The models are absorbed by the same hydrogen column density, integrated for infinite exposure and folded through the pn response.}
    \label{fig:powerlaw_apec}
\end{figure}

The model's reliance on the hard-band is further justified in by spectral degeneracies present in the soft X-ray regime (\autoref{fig:powerlaw_apec}). Despite the fact that the 0.5-2 keV band contains approximately 100 times more photons than that at energies greater than 2.5 keV, the 3D-GradCAM importance is higher in the latter for the cluster class. 

\subsection{Limitations}
While joint spatio-spectral representations improve performance, supervised models remain bounded by their training distributions. Deploying this pipeline across real observations or alternative X-ray observatories introduces key systematic considerations.

Within our current framework, variations in detector response matrices (RMFs) and exposure times alter raw photon counts. Standardising training inputs to exposure-normalised count rates ($\text{cts s}^{-1} \text{ keV}^{-1} \text{ arcsec}^{-2}$) and focussing on EPIC-PN, which dominates XMM statistics, provides a stable, exposure-independent representation. Furthermore, our current baseline assumes near-on-axis targets ($< 3'$). Real survey detections span the full field of view, where the PSF suffers from off-axis distortion and vignetting. Future iterations will address off-axis transferability by either parametrising off-axis angle directly as a conditional input during training, or employing Domain-Adversarial Neural Networks (DANNs; \citealt{ganin2016domain}) to enforce extraction of PSF-invariant features.

Beyond XMM-Newton, instrumental resolution varies significantly across X-ray observatories, from Chandra ($\sim 0.5''$) to XMM-Newton ($\sim 6''$)and eROSITA ($5.7\text{--}9.3''$) for on-axis FWHM.
Models trained on XMM-Newton cannot be applied zero-shot to other instruments without re-binning spatial grids or retraining on mission-specific PSFs.

Variations in energy bandpass and effective area profiles dictate how the models leverage spectral information. Our Grad-CAM interpretability analysis (Figure \ref{fig:gradcam1dz}) demonstrates that the network draws significant attention from the 6-10\,keV band to distinguish the high-energy exponential cut-off of thermal intracluster plasma from non-thermal AGN power-law emission. For soft-band optimized instruments like eROSITA, where effective area drops sharply above 2.5\,keV, an adapted model would need to rely primarily on soft-band spectral curvature and spatial morphology, where broader PSF cores make morphological separation more challenging. Conversely, future facilities such as the NewAthena Wide Field Imager (WFI; \citealt{nandra2013hot}), which combines hard-band sensitivity with sharp spatial resolution ($\sim$5''), present an ideal regime where joint 2D+1D and 3D spatio-spectral representations should remain highly effective.

\section{Conclusions}\label{sec:conclusions}

In this study we evaluated the efficacy of convolutional neural networks in discriminating galaxy clusters and AGN across 4 distinct X-ray data representations - 1D energy spectra, 2D imaging, combined spectra and imaging (2D+1D) and 3D binned event voxels (event cubes). Our results demonstrate that convolutional neural networks perform exceptionally at this task, obtaining high scores across all performance metrics. Notably, the models exhibit robust generalisation despite the inherent class imbalance of AGN to clusters,  suggesting that the learned features are sufficiently discriminative to overcome biased sampling. 

While 2D spatial information proved superior to 1D spectral data alone - aligning with traditional astronomical classification paradigms, the integration of both spatial and spectral properties yield a significant gain in diagnostic power.  The 3D model marginally outperformed the 2D+1D model, across most metrics. While the 2D+1D model achieved slightly higher ROC and PR curve AUC scores, the 3D model demonstrated superior consistency across the 9 cross validation runs. This indicates that the 3D kernels are more effective at capturing the intrinsic, multi-dimensional correlations within the event cubes, consistent with previous demonstrations in spatio-temporal learning tasks \citep{tran2015learning}. In contrast, factorised 2D+1D architectures may decouple these correlations, potentially leading to greater sensitivity to classification thresholds in our experiments.

A significant advantage of the 3D event cube approach is its ability to process raw counts directly, bypassing the traditional requirement for explicit background modelling. In 1D spectral analysis, classification often relies on background-subtracted products, this is non-trivial and can introduce systematic uncertainties and require manual calibration \citep{nevalainen2005xmm, ehlert2022probabilistic}. None the less, it would be possible for a CNN to classify the total spectra only. In contrast, our 3D model operates on total photon counts within the spatio-spectral voxels.

Quantitative interpretability analysis using 3D Grad-CAM reveals that the CNN has autonomously learned the fundamental morphological and spectral distinctions between Galaxy Clusters and AGN. Classification of Clusters is driven by a localised spectral focus in the very soft X-ray band (< 1 keV) as well as a broader influence from the hard X-ray band and a spatially centralised attention profile that captures the bulk of the extended intracluster medium. In contrast, AGN identification relies on a broad spectral scanning strategy centred in the soft regime and a spatially concentrated peak. This suggests the model distinguishes AGN by the high-signal-to-noise ratio of the central Gaussian-like PSF profile.

Our interpretability results reveal that the 3D CNN autonomously develops a dual-diagnostic strategy. By learning the relationship between a source's spatial envelope and the stable instrumental background, the 3D CNN effectively utilises stationary hard-band features as a reference. The ability to leverage the diffuse morphological signature against the instrumental background without manual subtraction demonstrates an end-to-end processing capability. Ultimately, this architecture streamlines the analysis pipeline and minimises human bias in background estimation, presenting a robust framework for identifying low-surface-brightness structures in future large-scale X-ray surveys. 

A preliminary zero-shot evaluation on real XXL Survey observations highlights the presence of sim-to-real domain shift while confirming that 3D spatio-spectral representations retain superior discriminative capability over lower-dimensional baselines (\autoref{sec:realobs}). To bridge this domain gap, and address current limitations, future work will incorporate off-axis PSF modelling and domain adaptation evaluated on realistic wide-sky simulations, namely, projected X-ray light cones over several 10 deg$^2$ with self-consistent cluster and AGN modelling (Muñoz et al in prep). These steps are essential to validate the model's performance in complex observational environments before deployment on large-scale survey catalogues.

\begin{acknowledgements}
The proof of concept of the work presented in this paper was established during an engineer internship carried out by Mateo Bentura (IMT Atlantique) under the supervision of MP at CEA Saclay.
We acknowledge input from Gabriele Matzeu and EG acknowledges funding from the German Aerospace Center (DLR).  

The code developed for this study is publicly available on GitHub at \url{https://github.com/MaggieLieu/3D_X-ray_Cluster_Detection.git}.
\end{acknowledgements}

\bibliographystyle{aa} 
\bibliography{reference}
\begin{appendix}
\section{Model architectures}

\begin{minipage}{\textwidth}
    \begin{minipage}{0.48\textwidth}
        \captionof{table}{1D CNN model architecture.}
        \begin{tabular}{|l|c|r|}
        \hline
        Layer (type) & Output Shape & Param \# \\ \hline
        Input (InputLayer) & (None, 50, 2) & 0 \\ \hline
        Conv (Conv1D) & (None, 50, 32) & 224 \\ \hline
        AP (AveragePooling1D) & (None, 25, 32) & 0 \\ \hline
        Conv1 (Conv1D) & (None, 25, 64) & 6,208\\ \hline
        AP1 (AveragePooling1D) & (None, 13, 64) & 0 \\ \hline
        Conv2 (Conv1D) & (None, 13, 128) & 24,704 \\ \hline
        AP2 (AveragePooling1D) & (None, 7, 128) & 0 \\ \hline
        Flatten (Flatten) & (None, 896) & 0 \\ \hline
        Dense (Dense) & (None, 64) & 57,408 \\ \hline
        Output (Dense) & (None, 1) & 65 \\ \hline
        \multicolumn{3}{|l|}{Total params: 265,829 (1.01 MB)} \\
        \multicolumn{3}{|l|}{Trainable params: 88,609 (346.13 KB)} \\
        \multicolumn{3}{|l|}{Non-trainable params: 0 (0.00 B)} \\ 
        \multicolumn{3}{|l|}{Optimizer params: 177,220 (692.27 KB)} \\ \hline
        \end{tabular}
        \label{tab:1d_cnn_model_summary}
    \end{minipage}
    \hfill
    \begin{minipage}{0.48\textwidth}
        \captionof{table}{2D CNN model architecture.}
        \begin{tabular}{|l|c|r|}
        \hline
        Layer (type) & Output Shape & Param \# \\ \hline
        Input (InputLayer) & (None, 32, 32, 1) & 0 \\ \hline
        Conv (Conv2D) & (None, 32, 32, 32)  & 320 \\ \hline
        AP (MaxPooling2D) & (None, 16, 16, 32)  & 0 \\ \hline
        Conv1 (Conv2D) & (None, 16, 16, 64) & 18,496 \\ \hline
        AP1 (MaxPooling2D) &  (None, 8, 8, 64)  & 0 \\ \hline
        Conv2 (Conv2D) & (None, 8, 8, 128) & 73,856 \\ \hline
        AP2 (MaxPooling2D) & (None, 4, 4, 128)  & 0 \\ \hline
        flatten (Flatten) &  (None, 2048)  & 0 \\ \hline
        Dense (Dense) & (None, 64)  &  131,136 \\ \hline
        Output (Dense) & (None, 1)   &  65 \\ \hline
        \multicolumn{3}{|l|}{Total params: 671,621 (2.56 MB)} \\
        \multicolumn{3}{|l|}{Trainable params: 223,873 (874.50 KB)} \\
        \multicolumn{3}{|l|}{Non-trainable params: 0 (0.00 B)} \\ 
        \multicolumn{3}{|l|}{Optimizer params: 447,748 (1.71 MB)} \\ \hline
        \end{tabular}
        \label{tab:2d_model_summary}
    \end{minipage}
    \begin{minipage}{\textwidth}
        \vspace{1cm}
        \captionof{table}{2D+1D CNN model architecture.}
        \centering
        \begin{tabular}{|l|l|r|l|}
        \hline
        Layer (type) & Output Shape & Param \# & Connected to \\ \hline
        Input2D (InputLayer) & (None, 32, 32, 1) & 0 & - \\ \hline
        Conv2D (Conv2D) & (None, 32, 32, 32) & 320 & Input2D[0][0] \\ \hline
        AP2D (AveragePooling2D) & (None, 16, 16, 32) & 0 & Conv2D[0][0] \\ \hline
        Conv2D\_1 (Conv2D) & (None, 16, 16, 64) & 18,496 & AP2D[0][0] \\ \hline
        AP2D\_1 (AveragePooling2D) & (None, 8, 8, 64) & 0 & Conv2D\_1[0][0] \\ \hline
        Conv2D\_2 (Conv2D) & (None, 8, 8, 128) &  73,856  & AP2D\_1[0][0] \\ \hline 
        AP2D\_2 (AveragePooling2D) & (None, 4, 4, 128) & 0 & Conv2D\_2[0][0] \\ \hline 
        Flatten2D (Flatten) & (None, 2048) & 0 & AP2D\_2[0][0] \\ \hline \hline
        Input1D (InputLayer) & (None, 50, 2) & 0 & - \\ \hline
        Conv1D (Conv1D) & (None, 50, 32) & 224 & Input1D[0][0] \\ \hline
        AP1D (AveragePooling1D) & (None, 25, 32)  & 0 & Conv1D[0][0] \\ \hline
        Conv1D\_1 (Conv1D) & (None, 25, 64) & 6,208 & AP1D[0][0] \\ \hline
        AP1D\_1 (AveragePooling1D) & (None, 13, 64)  & 0 & Conv1D\_1[0][0] \\ \hline
        Conv1D\_2 (Conv1D) & (None, 13, 128) & 24,704 & AP1D\_1[0][0] \\ \hline
        AP1D\_2 (AveragePooling1D) & (None, 7, 128)  & 0 & Conv1D\_2[0][0] \\ \hline
        Flatten1D (Flatten) & (None, 896) & 0 & AP1D\_2[0][0] \\ \hline
        \hline
        Concatenate (Concatenate) & (None, 2944) & 0 & \begin{tabular}[c]{@{}l@{}}Flatten2D[0][0],\\ Flatten1D[0][0]\end{tabular} \\ \hline
        Dense (Dense) & (None, 64) & 188,480 & Concatenate[0][0] \\ \hline
        Output (Dense) & (None, 1) & 65 & Dense[0][0] \\ \hline
        \multicolumn{4}{|l|}{Total params: 937,061 (3.57 MB)} \\
        \multicolumn{4}{|l|}{Trainable params: 312,353 (1.19 MB)} \\
        \multicolumn{4}{|l|}{Non-trainable params: 0 (0.00 B) } \\
        \multicolumn{4}{|l|}{Optimizer params: 624,708 (2.38 MB)} \\ \hline 
        \end{tabular}
        \label{tab:2dp1d_model_summary}
    \end{minipage}
\end{minipage}

\clearpage
\begin{minipage}{\textwidth}
    \centering
    \captionof{table}{3D CNN model architecture.}
    \begin{tabular}{|l|l|r|}
    \hline
    Layer (type) & Output Shape & Param \# \\ 
    \hline 
    Input (InputLayer)         & (None, 32, 32, 50, 1) & 0 \\ \hline
    Conv3D (Conv3D)          & (None, 32, 32, 50, 32) & 896 \\ \hline
    AP3D (AveragePooling3D) & (None, 16, 16, 25, 32) & 0 \\ \hline
    Conv3D\_1 (Conv3D)          & (None, 16, 16, 25, 64) & 55,360 \\ \hline
    AP3D\_1 (AveragePooling3D) & (None, 8, 8, 13, 64) & 0 \\ \hline
    Conv3D\_2 (Conv3D)          & (None, 8, 8, 13, 128) & 221,312 \\ \hline
    AP3D\_2 (AveragePooling3D) & (None, 4, 4, 7, 128) & 0 \\ \hline
    Flatten (Flatten)          & (None, 14336) & 0 \\ \hline
    Dense (Dense)            & (None, 128) & 1,835,136 \\ \hline
    Dense\_1 (Dense)            & (None, 64) & 8,256 \\ \hline
    Output (Dense)             & (None, 3) & 195 \\\hline 
    \multicolumn{3}{|l|}{Total params: 2,121,155 (8.09 MB)} \\
    \multicolumn{3}{|l|}{Trainable params: 2,121,155 (8.09 MB)} \\
    \multicolumn{3}{|l|}{Non-trainable params: 0 (0.00 B)} \\ \hline
    \end{tabular}
    \label{tab:3d_model_summary}

\noindent

\section{ROC and PR curves}

    \centering
    \includegraphics[width=0.49\linewidth]{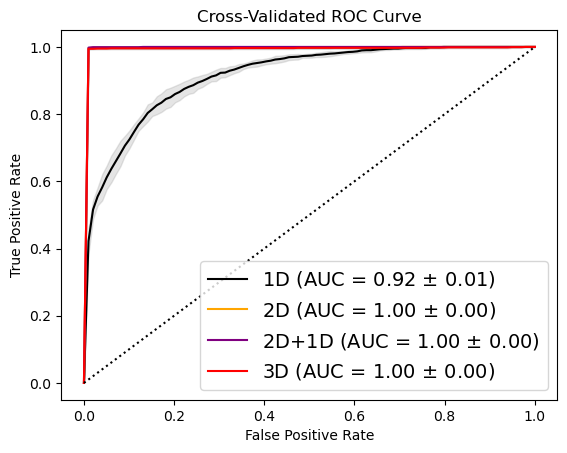}
    \includegraphics[width=0.49\linewidth]{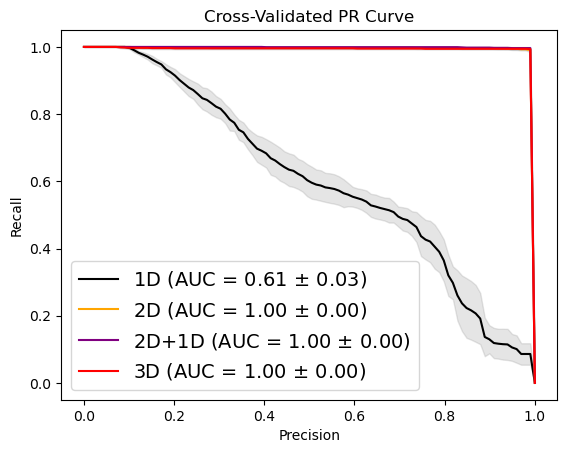}
    \captionof{figure}{Left: ROC curve with AUC scores. Right: Precision-recall plot.}
    \label{fig:ROC_PR}
\end{minipage}
\bigskip

\clearpage
\section{Multi-layer perceptron}\label{sec:mlp}
Traditionally, X-ray spectral data (1D data) is analysed on coarse binning schemes (e.g. 5 total and 5 background counts). Given the limited complexity, our initial approach employed a simple multi-layer perceptron (MLP) as the 1D model. MLPs are well-suited for low dimensional feature sets due to their ability to map non-linear relationships between small numbers of input features and output classes. This architecture, summarised in \autoref{tab:1d_model_summary}, provides sufficient capacity to capture the relevant features from the 1D spectral data while maintaining computational efficiency, a strategy previously shown to excel at X-ray spectral fitting and modelling \citep{matzeu2022new, parker2022agn}. We also explored a 3D MLP approach, treating the data as vectors of spatial ($x, y$) and energy ($E$) coordinates. However, since the number of photon counts per observation varies significantly, the required padding introduced substantial sparsity, hindering the model's ability to generalise. 

\begin{figure}
    \centering
    \includegraphics[width=\linewidth]{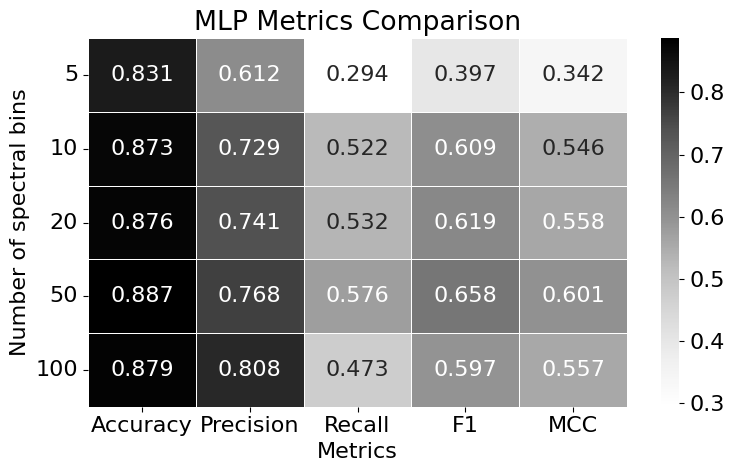}
    \caption{Using more spectral bins improves the 1D MLP model}
    \label{fig:spec bin model}
\end{figure}

\begin{figure}
    \centering
    \includegraphics[width=\linewidth]{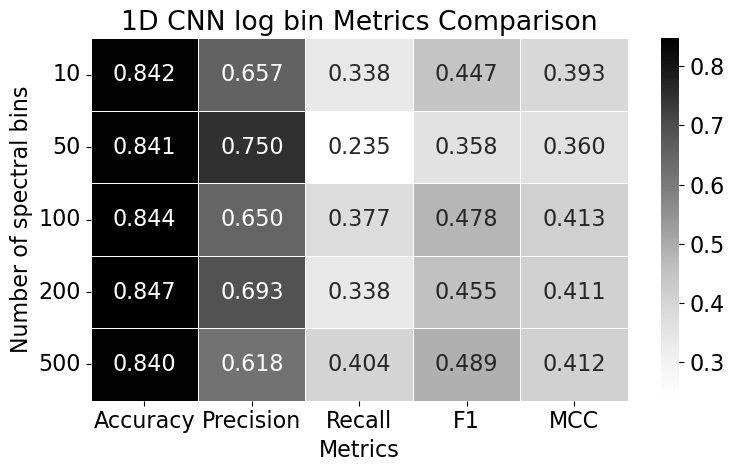}
    \caption{Using log spaced spectral bins between 0.5-10 keV in the 1D CNN model produces reduced performance and its less clear that increasing the number of spectral bins improves performance.}
    \label{fig:spec log bin cnn model}
\end{figure}

Further investigation, reveals that model performance improves with increasing spectral resolution \autoref{fig:spec bin model}. As the number of bins increases to 50, the data gains structural locality that a global MLP cannot exploit as efficiently as a convolutional architecture. Consequently, we transitioned to a CNN 1D model with 50 spectral bins \autoref{fig:spec bin  cnn model}, to better leverage these spatially correlated features and higher resolution spectra. Notably, the binning scheme is linearly spaced as log spaced bins produced inferior results (\autoref{fig:spec log bin cnn model}). The specific resolution was chosen to maintain consistency with the 3D CNN model where higher dimensionality would have incurred prohibitive computational costs and memory requirements. 

We additionally explored a 3D MLP architecture that processes the raw event list photons however, this requires padding for data lacking sufficient photons and computationally efficiency was poor. 

\begin{table}[hbp]
\caption{1D Model architecture for the 5 bin spectral input.}
\begin{tabular}{|l|c|r|}
\hline
Layer (type) & Output Shape & Param \# \\ \hline
Input (InputLayer) & (None, 5, 2) & 0 \\ \hline
Dense (Dense) & (None, 5, 32) & 96 \\ \hline
flatten (Flatten) & (None, 160) & 0 \\ \hline
Dense0 (Dense) & (None, 64) & 10,340 \\ \hline
Dense1 (Dense) & (None, 128) & 8,320 \\ \hline
Dense2 (Dense) & (None, 64) & 8,256 \\ \hline
Output (Dense) & (None, 1) & 65 \\ \hline
\multicolumn{3}{|l|}{Total params: 27,041 (105.63 KB)} \\
\multicolumn{3}{|l|}{Trainable params: 27,041 (105.63 KB)} \\
\multicolumn{3}{|l|}{Non-trainable params: 0 (0.00 B)} \\ \hline
\end{tabular}
\label{tab:1d_model_summary}
\end{table}

\begin{figure}
    \centering
    \includegraphics[width=\linewidth]{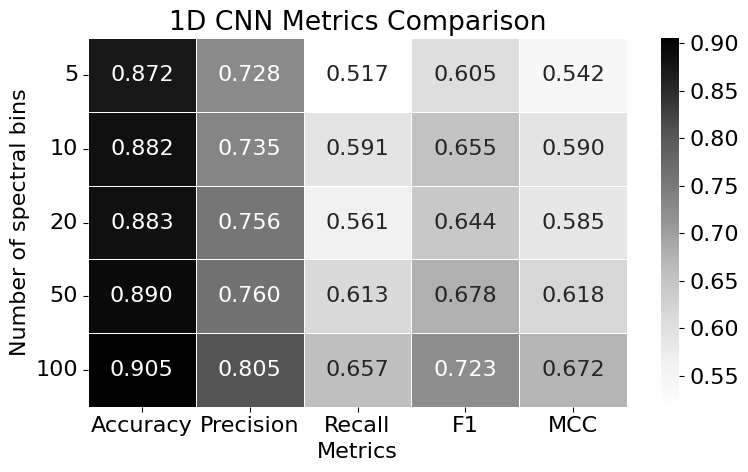}
    \caption{Using more spectral bins allows us to move back to 1D CNN model which improves the scoring.}
    \label{fig:spec bin cnn model}
\end{figure}

\section{Applications to real observations}\label{sec:realobs}
\begin{table*}[h]
\centering
\small
\caption{Model performance metrics on real observations.}
\label{tab:results_real}
\begin{tabular}{lccccccccccc}
\toprule
Model & TP & FP & FN & TN & Accuracy & Precision & Recall & F1 & MCC & ROC AUC & PR AUC \\
\midrule
2D+1D & 0 & 9 & 2 & 277 & 0.9618 & 0.0000 & 0.0000 & 0.0000 & 0.000 & 0.7675 & 0.0142 \\
3D & 1 & 16 & 1 & 270 & 0.9410 & 0.0588 & 0.5000 & 0.1053 & 0.1565 & 0.9073 & 0.0511
\end{tabular}
\end{table*}
To assess the transferability of our simulation-trained architectures to real observational data, we perform a zero-shot evaluation on a sample of confirmed X-ray sources from the XXL Survey \citep{pierre2016xxl, adami2018xxl, chiappetti2018xxl}. To ensure consistency with the parameter space of our training set, we select sources located within $3'$ of the boresight with net source counts between 40 and 60 photons, using the standard calibrated event lists from the XMM-Newton Pipeline Processing System (PPS).

After excluding five observations lacking calibrated files (ObsIDs 677680137, 677760132, 604873435, 604873436, 604873437), the resulting selection consists of 2 confirmed galaxy clusters and 286 confirmed AGNs. Due to the limited sample size training complex convolutional neural networks from scratch is beyond the scope of this paper. We therefore apply the simulation trained 2D+1D and 3D models directly as an out-of-distribution benchmark.

The classification results are presented in Table \ref{tab:results_real}. We note that while the XXL catalogue classifications serve as our benchmark ground truth, survey labels in the low count regime are not guaranteed to be infallible. Source blending, optical spectroscopic misidentification, and Poisson noise in spatial-extent fitting can introduce label contamination. Consequently, disagreements between model predictions and catalogue classifications do not strictly indicate network misclassification alone. To investigate the false-positive cluster classifications, we visually inspect archival imaging for optical and mid-infrared counterparts using the Dark Energy Survey Data Release 2 \citep[DES DR2][]{abbott2021dark} and the Wide-field Infrared Survey Explorer \citep[WISE][]{wright2010wide}. This visual inspection reveals several plausible uncatalogued candidates:
\begin{itemize}
    \item 3XLSS J233401.1-554900 shows clear optical overdensities consistent with a galaxy cluster candidate.
    \item 3XLSS J232951.5-560735 is potentially associated with cluster substructure
    \item 3XLSS J232725.0-535445 appears as a distant candidate with no detectable emission in WISE
    \item 3XLSS J232243.3-530814 is consistent with a compact galaxy group.
    \item 3XLSS J234224.4-550012 features four red galaxies $\sim 20"$ to the East accompanied by a marginal source concentration in WISE.
\end{itemize}
We emphasise however that 40-60 net photons represents a very low signal-to-noise regime. For extended cluster emission this flux range typically corresponds to high redshift systems ($z$ > 1) that lies beyond the photometric depth limit of DES. 
Furthermore the drop in performance compared to validation metrics on synthetic test sets highlights the challenge of sim-to-real domain shift. Real observations contain non-astrophysical artifacts that were omitted from our baseline simulations (e.g. residual soft proton background flaring, CCD chip gap boundaries and hot pixels.)

These findings demonstrated that while caution is required when applying simulation-trained models zero-shot to real observational domains, the 3D model remains more robust than the 2D+1D model under domain shift. This empirical validation confirms that joint spatio-spectral feature representations preserves essential physical signatures of the intracluster medium (ICM) and AGN power-law emission, while establishing a clear operational baseline for future domain-adaptation.

\end{appendix}
\end{document}